# The fate of Li and Be in stars and in the laboratory

V. Castellani[1,2,3] G. Fiorentini[4,5], B. Ricci[6,5] and O. Straniero [2,3].

[1] Dipartimento di Fisica dell'Università di Pisa, I-56100 Pisa.
[2] Osservatorio Astronomico di Collurania, I-64100 Teramo.
[3] Istituto Nazionale di Fisica Nucleare, LNGS, I-67010 Assergi.
[4] Dipartimento di Fisica dell'Università di Ferrara, I-44100 Ferrara.
[5] Istituto Nazionale di Fisica Nucleare, Sezione di Ferrara, I-44100 Ferrara.
[6] Scuola di Dottorato dell'Università di Padova, I-35100 Padova.
(May 22, 1995)

## Abstract

We connect the observed under-abundances of Li and Be in dwarfs, with recent results on nuclear cross sections at low energies: for collisions of protons with atomic or molecular targets, the measured cross sections seem too high with respect to extrapolations for bare nuclei. Phenomenologically, these anomalous nuclear interactions can be described in terms of an effective screening potential $U_{lab}$ in the range of few hundred eV: in the presence of the electron cloud, nuclei become more transparent to each other as if the effective collision energy is aumented by $U_{lab}$. This implies that fusion cross sections are enlarged and at the same time elastic cross sections are lowered. If something similar occurs in stellar plasma, the nuclear burning temperatures are lowered, whereas diffusion processes are enhanced. We find that the observed Li and Be abundances in the Hyades and in the Sun can be reproduced for effective screening potentials of the plasma in the range of 600-700 eV, close to that found by experiments in the laboratory.

*subject headings:* stars: abundance-nuclear reactions-diffusion.

# 1. INTRODUCTION

This paper is an attempt to connect two anomalies, one in stars and the other in the laboratory:

a) the abundances of Li and Be in dwarfs are generally too small with respect to the predictions of evolutionary calculations;

b) the measured nuclear cross sections at low energies for collisions of protons with atomic (or molecular) targets, $\sigma_{at}$, are too large, with respect to extrapolations of data taken at higher energies, where one can assume to have collisions of protons with bare nuclei, $\sigma_{nuc}$.

Concerning a) the problem is to account for the observed depletion of light elements in photospheric regions, the temperature - in these regions and in the convective layers below - being too low for nuclear burning, according to the nuclear reaction rates adopted in current evolutionary codes. Two well known evidences of this situation are given by the Sun and the Hyades, see Michaud & Charbonneau (1991) for a review of observational data and theoretical approaches.

We remark that the Li abundance problem is not alone, but has to be seen and explained in conjunction with the fate of other light elements, for example Be. The comparison between the present Sun and the meteoritic composition indicates that the solar photosphere is depleted by a factor about 100 for Li and about 2 for Be.

Concerning b), the situation is that for several nuclear reactions, at collision energies E in the $10 - 100$ KeV range the measured cross sections are too large with respect to expectation. If one writes

$$\sigma_{at}(E) = \sigma_{nuc}(E + U_{lab}) \quad , \tag{1}$$

one finds $U_{lab}$ much larger than the values $U_{el}$ given by theoretical calculations for the electron screening. For example, for collisions of protons with $^6$Li and $^7$Li nuclei one finds experimentally $U_{lab}^{p+Li} \approx 400$ eV (Engstler et al 1992), whereas $U_{el}^{p+Li} \approx 180$ eV (Bracci et al. 1990).

We call $U_{lab}$ an "effective screening potential in the laboratory" and we consider it as a parameter accounting for the (presently) inadequate theoretical treatment of nuclear reactions at very low energies. We remark that the condition $U_{lab} >> U_{el}$ is not an accident for just one reaction, but it seems to be a general problem, see e.g. Fiorentini, Kavanagh and Rolfs (1995).

Experiments in the laboratory, if correct, indicate that some effects due to the electron cloud around the colliding nuclei has been neglected. May be that also in stars we are neglecting something, connected with the interactions of the nuclei with the plasma. The aim of this paper is to investigate such a possibility.

As a working hypothesis, we thus make the following Ansatz: some anomaly also occurs in the stellar plasma, so that the nuclear cross sections in the plasma are given by

$$\sigma_{pl}(E) = \sigma_{nuc}(E + U_{pl}) \tag{2}$$

and we keep $U_{pl}$, the "effective plasma screening potential" for the reaction, as a free variable allowing it to be in the range of a few hundred eV, i.e. significantly larger than found in standard treatments of plasma screening (as an example, for collisions of two nuclei with charges $Z_1$ and $Z_2$ at the bottom of the solar convective region where the Debye length is $R_D \approx 2.\ 10^{-8}$cm, in the weak screening approximation one has $U_w = Z_1 Z_2 e^2 / R_D \approx 7 Z_1 Z_2 [\text{ eV}])$.

We also allow $U_{pl}$ to be specific of the nuclear reaction,

$$U_{pl} = U_{pl}^{p+^A Z} \quad , \tag{3}$$

but we assume that it is independent of plasma properties (at least in the region of interest to us), much in the same way as the observed $U_{lab}$ values are approximately independent of the chemical or physical state of the target.

Equation (2) implies that in stellar calculations the Maxwellian averaged burning rates at temperature T become:

$$\lambda_{pl}(T) = \lambda_{nuc}(T) exp(U_{pl}/kT) \tag{4}$$

where $\lambda_{nuc}$ are the burning rates for bare nuclei. This clearly gives significant changes in the temperature $T_{NB}$ at which nuclear burning becomes efficient (which we define from the condition $\lambda_{nuc}(T_{NB})t_\odot = 1$, $t_\odot$ being the Sun's age). For example, the burning temperature of $^7$Li in the Sun, usually quoted as $T_{NB}$=2.5 (here and in the following $T$ means the temperature in units of $10^6$ K), becomes $T_{NB}$=2.1 for $U_{pl}^{p+Li}$=700 eV, whereas for $^9$Be one moves from $T_{NB}$=3.5 to $T_{NB}$=3 passing from $U_{pl}^{p+Be}$=0 eV to $U_{pl}^{p+Be}$=700 eV.

The basic question addressed in this paper is thus the following: can we find a range of $U_{pl}$ values which can account for the observational data in stars?

Firstly, in section 2 we present a discussion of the observational data for 1M$_\odot$ stars, in the conventional framework, with the following main conclusions.
a) We exclude any attempt of explaining Be depletion in terms of mixing the photosphere with high temperature ($T \approx$3.5) regions where Be can be burnt, since Li would be completely destroyed.
b) As a consequence, Be has to be hidden (without being burnt) below the convective zone by diffusion processes.
c) From the comparison essentially among meteorites, Hyades and the present Sun, we can estimate Li depletion in the pre-main sequence (PMS) and in the main sequence (MS) phases. Li depletion occurs in <u>both</u> phases.
d) The same diffusive processes which hide Be are not sufficient to account for the Li depletion during the main sequence. Li has to be burnt also during the main sequence.

In section 3 we summarize, from measurements in the laboratory, the knowledge of $U_{lab}$ for the reactions of interest to us and we present some arguments for assuming that the plasma screening potential $U_{pl}$ is (roughly) independent of plasma properties, and similar (not necessarily equal) to $U_{lab}$.

We also show that introducing $U_{pl}$ significantly enhances the diffusion coefficients. The anomalously large measured nuclear cross sections seem to indicate that nuclei are more transparent to each other and this suggests that elastic cross sections are lowered, so that diffusion can be enhanced.

In section 4 we investigate Li burning in the pre-main sequence and in the early main sequence phases. The study of a young open cluster such as the Hyades, for which many observational data are available, is of particular interest since one can compare stars with different masses while the other parameters (age, chemical composition...) are the same. Stars with different masses have different values of temperature at the bottom of the convective zone, stars with lower surface temperature being hotter at the bottom of the convective zone. In this way one can essentially explore the burning rate as a function of temperature. In a sense, this is the astrophysical equivalent of an experiment in the laboratory where the energy of the accelerated particles is varied so as to measure the energy dependence of nuclear cross sections. We will find that for $U_{pl}^{p+Li} \approx 600 - 700 eV$ one can account for the observational data on the Hyades.

In section 5 we discuss Li (and Be) depletion of the convective zone in the main sequence phase of the Sun. This depletion can be due to several processes:
a) burning at the bottom of the convective zone, which clearly gets easier if the anomalous burning rates of equation (4) are considered;
b) concentration driven diffusion: for Li, the presence of an anomalous screening potential results in a steep rise of the burning rate just below the convective zone; this induces a strong Li concentration gradient which can drive diffusion;
c) gravitational and thermal diffusion.

The efficiency of processes b) and c) depends on the values of the diffusion coefficients. With the help of the observed value for the Be depletion in the Sun, which we consistently assume to be due to diffusive processes, we will fix the diffusion coefficients for Li and we will study the effects of processes a)-c). We will show that again for $U_{pl}^{p+Li} \approx$ 600-700 eV we have a mechanism which can account for the observed solar Li depletion.

Throughout all the discussion, the properties of the bottom of the convective region play a key role. Within a phenomenological approach, these properties will be determined from observational (helioseismoligical) data whenever possible, otherwise we will use results which are common to most solar model calculations. All this information is summarized in the Appendix B, whereas Appendix A contains a short outline of the evolutionary code FRANEC (Frascati Raphson Newton Evolutionary Code) we are using.

In conclusion, the range of $U_{pl}^{p+Li}$ which is needed for solar Li depletion in the main sequence overlaps with that reproducing the Li abundances in the Hyades and it is not far from the value of $U_{lab}^{p+Li}$ measured in the laboratory. A short discussion of future prospects is given in the final section.

## 2. OBSERVATIONAL DATA ON LI AND BE DEPLETION IN 1M$_\odot$ STARS

Observational data on Li depletion in different astrophysical contexts have been recently reviewed by Michaud & Charbonneau (1991). From the available data we selected those pertaining to 1M$_\odot$ stars in various stellar clusters. Assuming that the selected stars have similar chemical composition, the graph shown in Fig. 1 tells us the history of photospheric Li abundance in 1M$_\odot$ star. The very stable structure of main sequence stars implies a constant efficiency of physical mechanisms and thus suggests a linear time dependence of photospheric abundances. Indeed, the data shown in Fig. 1, are well fitted by a straight line,

$$y_{Li}(t) = (2.35 \pm 0.2) - (0.27 \pm 0.05)t \quad , \tag{5}$$

where as usual $y_a = 12 + log(N_a/N_H)$, $t$ is the age in Gy and the quoted errors are to be taken as indicative of the uncertainties of the observational data. There is a clear indication that some Litihum depletion occurs during the main sequence. For the solar age we get:

$$\Delta y_{Li}(\text{MS}) = 1.2 \pm 0.25 \tag{6}$$

In Fig.1 we have also shown the meteoritic value, which tentatively can be taken as representative of the initial composition. The linear fit, eq. (5) then implies that a significant Li depletion also occurred in the pre-main sequence phase:

$$\Delta y_{Li}(\text{PMS}) = 1.0 \pm 0.2 \tag{7}$$

We note that this result is consistent with observational data from TT tauri which for 1M$\odot$ stars give only an upper bound $\Delta y_{Li} < 1$ ) (Zappalà 1972; Magazzù, Rebolo and Pavlenko 1992).

In stellar environments we do not have so far any observation about the relative abundances of $^6$Li and $^7$Li. The burning rates in the region of interest to us ($T$=2-4) can be espressed as (see Fig. 2):

$$\lambda_{nuc}^{p+^AZ} = B_{AZ}\rho X T^{\alpha_{AZ}} \quad , \tag{8}$$

with:

$$B_{^6Li} = 9.5 \cdot 10^{-7} Gy^{-1} \; ; \quad \alpha_{^6Li} = 19.6 \tag{9a}$$

$$B_{^7Li} = 1.2 \cdot 10^{-8} Gy^{-1} \; ; \quad \alpha_{^7Li} = 19.7, \tag{9b}$$

X being the Hydrogen mass fraction and $\rho$ the density, here and in the following in $g/cm^3$.

The burning rate of $^6Li$ is thus about a factor hundred higher than that of $^7Li$. The small initial $^6Li$ fraction (about 8% in meteorites), is thus completely destroyed when (and if) $^7Li$ is burnt. For this reason in the following we will neglect $^6Li$ and refer for brevity to Li when actually discussing $^7Li$ abundance.

The few available data about Be in 1M$_\odot$ stars are summarized in Table 1. Briefly, the solar abundance is one-halfof that in the meteorites, and data from Hyades for 1M$_\odot$ stars cannot discriminate between the solar and the meteoritic values, at present. Anyhow, this is important to tell the fate of Be. In the temperature region of interest to us the conventional ($U_{pl}$=0) burning rate can be expressed again using equation (8). When the two final channels $\alpha +^6 Li$ and d+2$\alpha$ are summed one has:

$$B_{^9Be} = 2.5 \cdot 10^{-13} Gy^{-1} \; ; \quad \alpha_{^9Be} = 23.5 \tag{10}$$

Thus the ratio between the burning rates of $^7$Li and $^9$Be is

$$R = \lambda_{nuc}^{p+^7Li}/\lambda_{nuc}^{p+^9Be} = 4.8 \cdot 10^4 \quad T^{-3.8} \tag{11}$$

Li and Be nuclei are close in mass and charge (also in any calculation the diffusion coefficients come out to be very similar for both nuclei) and it is natural to assume that any mixing mechanism is the same for both nuclei, so that they will experience the same temperatures. It is then clear from equation (11), see also Fig. 2, that at temperatures such that Be can be burnt ($T \approx 3.5$) Li would be completely destroyed!!

This rules out any attempt to explain Be depletion in terms of mixing the photosphere with high temperature regions where Be can be burnt.

There are only two possile outcomes:
a) Be can be burnt at significantly lower temperatures, which are not too dangerous for Li survival.
b) Be depletion is just a diffusion process: Be is hidden (not burnt) below the convective zone.

As will be discussed in sections 4 and 5, Be could be burnt in $1M_\odot$ stars for $U_{pl}^{p+Be} \approx 1700$ eV, a value which seems too high, although it is not excluded by the few experimental laboratory data presently available, see next section. We are thus left with the conclusion that Be is hidden just below the convective zone by diffusion. This requires the long times of the main sequence phase, and is due to pressure and thermal diffusion, concentration gradients being too low at the bottom of the convective zone in order to affect the Be diffusion.

The same processes will also affect the fate of Li, due to the similarity of the two elements. We thus conclude that during the main sequence the Sun has lost a fraction

$$\Delta y_{Li}^{PT}(\text{MS}) \approx \Delta y_{Be} = 0.27 \pm 0.10 \qquad (12)$$

where the index PT reminds the effects of pressure and temperature. The comparison between equation (6) and equation (12) shows that other effects are to be added, in order to account for the solar Li depletion during the MS phase.

Before closing this section, let us summarize the phenomenological picture we presented, recalling also the results of current evolutionary calculations.
i)There is observational evidence for a significant Li burning in the pre-main sequence phase of a $1M_\odot$ star. However, calculations in the standard framework (e.g. Proffitt & Michaud 1989) do not achieve significant Li burning, essentially since the star is not hot enough during the short pre-main sequence times. Either the pre-main sequence models are wrong or Li can be burnt more easily than believed.
ii) During the main sequence, Li diffusion due to PT effects, as estimated from data on Be, accounts just for a fraction of the Li depletion, and thus Li burning has to occur also in the main sequence. From helioseismology, and in agreement with several evolutionary calculations (see Appendix B), the temperature at the bottom of the convective zone of the Sun is determined to be in the range $T_b=2.1$-$2.3$ and this temperature should not have changed by more than 10% during the previous Sun evolution. This leads again to the conclusion that L can be burnt at temperatures smaller than currently believed.
iii)We ascribe Be depletion in the Sun to diffusion processes. As is well known (see section 5 and Appendix B), theoretical calculations (neglecting turbulence) yield a too small Be depletion.

A mechanism that at the same time would give lower burning temperatures and faster diffusion than in the standard framework would be welcome.

# 3. NUCLEAR BURNING IN THE LABORATORY: EXPERIMENTAL RESULTS AND SUGGESTIONS FOR STELLAR INTERIORS

An extensive experimental investigation of nuclear reactions between charged particles has been performed at very low collision energies (E≈10-100 KeV) in the last few years, particularly by the Rolfs group (Assenbaum et al. 1987; Engstler et al. 1992 and references therein).

The striking result is that at these low energies the nuclear cross sections are generally larger then expected by extrapolating data taken at higher energies, where the effect of the electron cloud surrounding the target nuclei can be neglected. By parametrizing the cross section as in equation (1), one extracts from the experiments the values reported in Table 2 for the effective screening potential $U_{lab}$.

Electron screening effects have been calculated by using different approximations: adiabatic/sudden (Bracci et al. 1990), classical trajectory Monte Carlo method (Bracci et al. 1989) and in a few cases with an ab initio quantum mechanical dynamical calculation (Bracci et al. 1991). The maximal value of the electron screening potential is obtained in the adiabatic limit, which corresponds to the maximal energy which, consistently with quantum mechanics, can be transferred from the electrons to the nuclear motion (Bracci et al 1989). In this limit, for collisions between two atomic systems A and B yielding at zero internuclear distance the compound atomic system A+B, the screening potential is given by:

$$U_{el}^{A+B} = E(A+B) - E(A) - E(B) \quad , \tag{13}$$

where E(i) are the atomic binding energies.

Although the experimental errors are individually large, the measured values $U_{lab}$ are systematically larger then $U_{el}$, see again Table 2. The reason for this discrepancy is not understood. We take the attitude that it is due to some inadequacy of the theoretical treatment and try to learn directly from experiments the properties of $U_{lab}$.

Reactions between H isotopes and Li isotopes are of particular interest for the present discussion. As one sees from Table 2, the values of $U_{lab}^{H+Li}$ are, whithin errors, i) independent of the isotopes which are interacting and ii) independent of the chemical state of the target. As remarked in Engstler et al. (1992), the first point suggests that the problem is not due to nuclear physics but it is related to interaction with the electron cloud. On the other hand, the second point shows that the effect is weakly sensitive to the detailed structure of the electron cloud.

Concerning $^9$Be(p,$\alpha$)$^6$Li and $^9$Be(p,d)2$\alpha$ reactions, the only published data at low enough energies are from Sierk & Tombrello (1973). At the four lowest measured energies, data show an increase of the astrophysical factor corresponding to $U_{lab}$ in the range of about 1 KeV, however the statistics is too poor for getting a definite answer. Preliminary results from Rolfs group (Zahnow et al. 1994) seem not to confirm the low energy enhancement.

In a stellar plasma of interest to us ($T$=2-4, $\rho \approx (0.1-1)$, $X \approx 0.7$) only a small fraction of Li or Be atoms are not dissociated. On the other hand, it is worth observing that the plasma cloud around a Li nucleus is similar to the electron cloud around the target Li nucleus in the laboratory, as in both cases the mean interparticle distance is $d \approx 10^{-8} cm$ and the electron velocity is $v_{el} \approx 10^8 cm/sec$. It is thus reasonable to assume an effective screening potential in the plasma $U_{pl}^{p+Li}$ of the same order of magnitude as that measured in the laboratory, in the range - say - of a few hundred eV.

In addition, $d$ and $v_{el}$ change weakly inside the plasma region defined above. Reminding the weak sensitivity of $U_{lab}^{H+Li}$ to the physical state of the target, it is thus natural to take $U_{pl}^{p+Li}$ as approximately independent of the plasma parameters.

Although the origin of the anomalous screening in the laboratory is unknown, if this phenomenon exists it should show up also in elastic collisions between nuclei, and thus it could affect the diffusion of nuclei through the plasma. This can be easily understood by observing that the anomalously large measured nuclear cross sections seem to indicate that nuclei are more transparent to each other. Actually, it is experimentally know that elastic cross sections of charghed nuclei are smaller than the Rutherford values at low energies (Huttel et al. 1985), however measurements of interst to stellar physics are so far not available.

Again with a phenomenological approach, we know from nuclear fusion measurements that the Coulomb barrier is lowered by an amount $U_{lab}$. Let us assume that a similar process occurs in the plasma, so that the effective interaction between two nuclei with charges $Z_1$ and $Z_2$ at distance $r$ is given by:

$$V(r) = Z_1 Z_2 e^2 / r - U_{pl}, \tag{14}$$

The relative nuclear motion is thus the same as in a pure Coulomb potential, the collision energy being increased by a quantity $U_{pl}$:

$$E \to E_{eff} = E + U_{pl} \tag{15}$$

The Coulomb cross section being $\sigma_{Cou}(E) = (Z_1 Z_2 e^2 / E)^2$, this means that in the plasma one has:

$$\sigma_{Cou,pl}(E) = \sigma_{Cou}(E + U_{pl}) = \left(\frac{Z_1 Z_2 e^2}{E + U_{pl}}\right)^2. \tag{16}$$

For a simple estimate of this effect we note that the diffusion coefficient D are inversely proportional to the Coulomb cross section. By replacing E$\to kT$ in eq. (16) one gets:

$$D(U_{pl}) \approx D(0)[1 + \frac{U_{pl}}{kT}]^2 \quad . \tag{17}$$

A more precise estimate, properly taking into account the energy distribution for the colliding particle and still neglecting variation of the Coulomb logarithm, gives:

$$D(U_{pl}) = D(0) \, / \, \int_0^\infty dx \frac{e^{-x} x^2}{(x + U_{pl}/kT)^2} \tag{18}$$

The result of the numerical evaluation of the integral is well approximated by the following expression:

$$D(U_{pl}) = D(0)[1 + 3\frac{U_{pl}}{kT} + \frac{1}{2}(\frac{U_{pl}}{kT})^2] \quad . \tag{19}$$

This means that, in the region of interest to us ($kT \approx 200$ eV), an effective screening potential $U_{pl}$ of few hundred eV corresponds to diffusion coefficients increased by an order of magnitude.

We do not attempt, here, to calculate all the diffusion coefficients relevant for the study of Li and Be. For this goal, at least one should have an estimate of the screening potential for the collisions of these ions with <u>both</u> H and He nuclei. It suffices here to remark that screening potentials of the same order as those measured in the laboratory can significantly enhance the rate of diffusion processes.

## 4. LI AND BE IN HYADES LOW MAIN SEQUENCE STARS

As already mentioned, surface abundances of light elements in low mass stars are the result of the concurrent action of two different physical processes: nuclear burning at the bottom of the convective envelope and atomic diffusion. It is generally difficult to discriminate between the different contributions; however, when stellar ages are short enough with respect to the diffusion timescale, we have the opportunity to study the burning reaction rates for different values of the temperature at the bottom of the convective envelope in stars with different masses. This is the case of the Hyades, for which rather accurate measurements of Li and Be abundances are available (Boesgaard & Tripicco 1986; Duncan & Jones 1983; Cayrel et al 1984; Thorburn et al. 1993; Boesgaard, Heacox and Conti 1977; Garcia Lopez et al. 1994). As is well known, although in the pre-main sequence the temperature at the bottom of the convective envelope reaches values larger then $T_b = 3.5$, the stellar lifetime is so short (less than 50 Myr) that classical stellar models cannot account for the observed Li depletion in the Hyades unevolved main sequence stars with $T_{eff} < 6000\ K$.

In Figs. 3-4 we report the surface abundance evolution of Li and Be for a $1 M_\odot$ star with solar chemical composition calculated for various values of the effective screening potential.

These results have been obtained by means of FRANEC (see Appendix A), neglecting diffusion. For all calculations presented in this section we use Z=0.02, Y=0.29 and the mixing length is $\alpha = 2.25$.

For $U_{pl}^{p+Li} = 0$ Li depletion is negligible, in agreement with previous well known results (Proffit & Micheaud 1989, D'Antona & Mazzitelli 1984) but in contrast to observed abundances. Li abundance depletion in agreement with the observational result (i.e. $\Delta y_{Li} \approx 1$) can be obtained for $U_{pl}^{p+Li} \approx (500 - 700)$ eV.

Note that most of the variation of the Li abundance occurs in the pre-main sequence phase, a really negligible variation taking place between the end of pre-main sequence and the estimated age of this cluster ($t_{Hy} = 0.8 \pm 0.2$ Gy, see e.g. Castellani, Chieffi and Straniero 1992), as the bottom of the convective region moves towards more external and cooler regions. In this context, Hyades abundances are representative of the pre-main sequences nucleosynthesis.

An ad hoc effective screening potential could obviously reduce the amount of surface Li to the observed value in a star of a given mass. However it is not clear that the same value of $U_{pl}^{p+Li}$ can be adequate for stars of different masses. That this is the case is shown in Fig 5, where we present the Li abundance for low mass Hyades as a function of their effective temperature. A value of $U_{pl}^{p+Li} \approx 700$ eV reproduces the trend of the observational data. It is worth noticing that for $U_{pl}^{p+Li}=0$ our results are again similar to those of other authors.

Let us remark that the temperature at the bottom of the convective envelope depends, for a given stellar mass, on the assumed chemical composition. A larger metallicity could imply a higher temperature at the bottom (for a fixed value of the He abundance). We do not attempt here a detailed discussion of this point, and we only note that the value of $U_{pl}^{p+Li}$ we just derived using Z=0.02 is, strictly speaking, an upper bound.

No significant Be depletion is observed for Hyades stars with low effective temperatures (see Garcia Lopez et al. 1994). In this case, available data can be used to set an upper limit to the effective screening potential for p+Be reaction. Fig. 6 indicates that the observed Be abundances allow $U_{pl}^{p+Be} \leq 1700$ eV .

## 5. FROM THE ZAMS TO THE PRESENT SUN

Within a phenomenological approach, without resorting to detailed evolutionary calculations, we study now the time evolution of the Li and Be abundances in the solar photosphere during the main sequence starting from Zero Age Main Sequence (ZAMS). Firstly we will write down the general equations for the evolution of heavy element abundances in the convective zone, showing that they are essentially determined from the properties of the innermost convective layer, essentially the temperature $T_b$, the distance from the solar center $R_b$, the density $\rho_b$, the pressure and temperature logarithmic derivatives, $dlnP/d(R/R_\odot)$ and $dlnT/d(R/R_\odot)$.

In the spirit expressed at the beginning of this section, we will assume that all these quantities change smoothly along the main sequence, their time dependence being of the form

$$O_\alpha = O_{\alpha\odot}[1 + \beta_\alpha(t - t_\odot)/t_\odot] \tag{20}$$

For the slopes $\beta_\alpha$ we will use results of our evolutionary code including diffusion, see column b of Table 3, and the values $O_{\odot\alpha}$ at the bottom of the convective zone of the present Sun are determined from observational (helioseismological) data whenever possible, otherwise we will use results which are common to several solar model calculations, all this information being summarized in the Appendix B.

In practice, to a very good approximation one could assume that all quantities, but the temperature $T_b$, are constant. Since the nuclear burning rates vary strongly with temperature, the time dependence of $T_b$ cannot be ignored and a few per cent decrease from ZAMS to the present Sun is important. We will take the temperature at the bottom of the convective layer in the present Sun, $T_{b\odot}$, as a parameter in the range:

$$T_{b\odot} = (2.1 - 2.3). \tag{21}$$

This interval is the result of inversion of helioseismological data (Christensen-Dalsgaard, Gough and Thompson 1991) and is consistent with evolutionary calculations where diffusion is considered, see again the Appendix B.

On these grounds we will calculate Li and Be abundances in a $1M_\odot$ star of solar chemical composition along the main sequence.

### 5.1 The basic equations

Due to the short characteristic times of the convection processes, the abundances $y_i$ ($y_i = 12 + log_{10}(N_i/N_H)$ and i=Li or Be) of each element in the photosphere are the same as in any point of the convective zone. The time evolution of $y_i$ can be represented as:

$$\frac{dy_i}{dt} = (log_{10}e)\frac{d(lnM_i)}{dt} = (log_{10}e)(- <\lambda_{pl}^{p+i}> + v_i\rho_i S/M_i) \qquad i =^7 Li, ^9 Be \qquad (22)$$

where $M_i$ is the mass of the i-th element contained in the convective zone, and the two terms on the right hand side, which we are going to define and to discuss in the following, represent respectively the contributions to the i-th element depletion due to the nuclear burning in the convective zone and due to the outflow towards the inner radiative region.

Coming to the first term, $<\lambda_{pl}^{p+i}>$ is the nuclear burning rate averaged over the convective zone

$$<\lambda_{pl}^{p+i}> = \frac{\int dM \lambda_{pl}^{p+i}}{\int dM}, \qquad (23)$$

where the integral is over the convective zone and the nuclear burning rates $\lambda_{pl}^{p+i}$ are given by equations (4) and (8).

Due to the strong temperature dependence, nuclear burning occurs most likely at the bottom of the convective zone, and its average rate is essentially related to the properties of the innermost layer. By numerical experiments (performed at different solar ages and by varying the position of the bottom of the convective zone) we find, to an accuracy of about 10%,

$$<\lambda_{pl}^{p+i}> = \frac{1}{8} \lambda_{pl}^{p+i}\bigg|_{bottom} = \frac{1}{8} B_i \rho_b X_b T_b^{\alpha_i} exp(\frac{U_{pl}^{p+i}}{kT_b}) \qquad (24)$$

and in the following we will fix $X_b = X_{b\odot} = 0.7$, $\rho_{b\odot} = 0.2$ g cm$^{-3}$.

Coming to the last term in equation (22), $v_i$ and $\rho_i$ are respectively the diffusion velocity and the density of the i-th element calculated at the bottom of the convective layer and $S = 4\pi R_b^2$ is the surface defining the border between the convective and the radiative region. We will take:

$$R_{b\odot} = 0.71 R_\odot \qquad (25)$$

Clearly the factor $\rho_i/M_i$ does not depend on the element one is considering, and one has:

$$\rho_i/M_i = \rho_b/M_{conv} \qquad (26)$$

where $M_{conv}$ is the mass contained in the convective zone; we will take $M_{conv\odot} = 0.025 M_\odot$.

Diffusion of the i-th element is driven by pressure gradients, thermal gradients and concentration gradients. At the bottom of the convective region the concentration gradients of H and He are negligible, and we only take into account Li (Be) concentration gradients, as these elements can be burnt in radiative regions which are closeby. We write thus for the diffusion velocity $v_i = \vec{v}_i \cdot \hat{R}$:

$$v_i = D_o[A_{C_i}\frac{dlnC_i}{d(R/R_\odot)} + A_P\frac{dlnP}{d(R/R_\odot)} + A_T\frac{dlnT}{d(R/R_\odot)}] \tag{27}$$

where the scale factor $D_o$ can be written as:

$$D_o = 1.2 \cdot 10^{-11} \ T^{5/2}/\rho_b \quad cm/sec \tag{28}$$

the density being measured in c.g.s units

The coefficients $A_P, A_T$ and $A_C$ are weakly dependent on temperature and density, at least in the region of interest to us, and we will take them as constant. The values for parameters representative of the present solar convective basis have been calculated following Thoul et al. (1994) and are shown in Table 4.

One notes that the Li and Be coefficients are similar and that for both elements the concentration coefficient is typically a factor 10 smaller than that of pressure. It has to be remarked, for future applications, that this is not an accident, but rather it is a consequence of the general relation between mobility and diffusion coefficients (see for example Landau & Lifshitz, §12 of Physical Kinetics and §59 of Fluid Mechanics). Pressure and thermal coefficients are comparable; however, as is well known, in the Sun thermal diffusion is less important than the pressure term, the thermal gradient being significantly smaller than that of pressure.

As discussed in section 3, the actual diffusion coefficients could differ from those calculated above due to the effects of screening, or to some turbulent process. For this reason, we will introduce a multiplicative parameter $\eta$ in front of equation (28):

$$D_o \to D = \eta D_o, \tag{29}$$

which we assume to be the same for Li and Be and which we will fix later from the observed Be depletion. Coming to the pressure and thermal gradients in equation (27), we note (see Appendix B) that for the present Sun all calculations give similar values at the bottom of the convective zone, close to:

$$\left.\frac{dlnP}{d(R/R_\odot)}\right|_\odot = -12 \tag{30a}$$

$$\left.\frac{dlnT}{d(R/R_\odot)}\right|_\odot = -5 \tag{30b}$$

We can approximately estimate the concentration gradient, by assuming that at any time the concentration below the bottom of the convective zone is determined by the nuclear burning rate only, i.e. as a first approximation we neglect diffusion since this process is slow in comparison with the others. In this way one gets:

$$C_i(t) = C_i(0) exp[-\int_0^t dt' \lambda_{pl}^{p+i}] \tag{31}$$

where time 0 corresponds to ZAMS. It follows that $dlnC_i/d(R/R_\odot)$ can be expressed in terms of the burning rate and temperature gradient at the bottom of the convective zone:

$$\frac{dlnC_i}{d(R/R\odot)} = -\int_0^t dt' \lambda_{pl}^{p+i}[\alpha_i - U_{pl}^{p+i}/kT]\frac{dlnT}{(dR/R_\odot)}. \tag{32}$$

In the approximation that the properties of the basis of the convective zone are unchanged during the main sequence, the above equation simplifies to:

$$\frac{dlnC_i}{d(R/R_\odot)} = -t\lambda_{pl}^{p+i}[\alpha_i - U_{pl}^{p+i}/kT]\frac{dlnT}{(dR/R_\odot)}. \tag{33}$$

### 5.2 Results

By integrating numerically equation (22) from ZAMS to the present Sun, we find the results shown in Figs. 7-10, which deserve several comments.

i) As expected, see Fig. 7 dashed curve, no significant Li depletion during the main sequence occurs, for the conventional ($U_{pl}^{p+Li} = 0$) nuclear burning rates and for standard ($\eta = 1$) diffusion coefficients.

ii) For the same value which reproduces the observed Li abundance in $1M_\odot$ Hyades stars, $U_{pl}^{p+Li} \approx 700$ eV, Li is burnt durning the main sequence but not enough to account for the present solar value. Standard diffusion terms ($\eta = 1$) contribute little to Li depletion, see Fig. 7.

iii) For large enough diffusion coefficients ($\eta = 5$), the Li depletion increases, in particular we note that the contribution of the concentration driven diffusion is relevant, see Fig. 8.

iv) For standard diffusion coefficients, the Be abundance is unchanged from the Hyades age to the present Sun, even for very large values of $U_{pl}^{p+Be}$, see Fig. 9 full curves. In other words, Be cannot be burnt. On the other hand, for $\eta \approx 5$ (dashed curves in Fig. 9) diffusion is effective to bring below the convective zone an amount in agreement with observational data. We remark that $\eta \approx 5$ corresponds to screening potentials of the order of few hundred eV, see section 3. We thus empirically choose $\eta = 5$, for both Be and Li.

v) It is worth observing, see Fig. 10, that for $\eta \approx 5$ and $U_{pl}^{p+Li} \approx 700$ eV the Li depletion during the main sequence is $\Delta y_{Li} \approx 1$ in agreement with observational data.

We have also evaluated the Li surface abundance evolution in a $1M_\odot$ star directly with a recent version of the FRANEC code where diffusion of He and heavy elements (C,N,O, Li and Be) is included. Diffusion coefficients have been calculated as in Thoule et al. (1994), a multiplicative factor $\eta = 5$ being included for Li and Be. The results are presented in Fig.

11, where one sees that again for $U_{pl}^{p+Li} \approx 600 - 700$ eV we are able to reproduce all the observational data from meteorites to the present Sun.

We remark that this range of $U_{pl}^{p+Li}$ corresponds to that needed for explaining the Li abundance in low main sequence Hyades, see section. 4 and it is not far from the value measured in the laboratory $U_{lab}^{p+Li} \approx 400$ eV.

## 6. CONCLUSIONS AND OUTLOOK

Briefly, the main result of this paper is the following: if one assumes that nuclei in the stellar plasma are more transparent to each other, as suggested by laboratory experiments in the 10-100 KeV region, one can understand light (Li and Be) elements abundances both in the Sun and in lower main sequence Hyades.

Quantitative results are summarized in Fig. 5, where we compare the calculated and the observed Li abundances in low main sequence Hyades, and in Fig. 11, where we show the time evolution of Li abundance in $1M_\odot$ stars, calculated by using our FRANEC code (element diffusion being included), together with the observational values.

In short, for effective plasma screening potentials of the order of few hundred eV the Li burning temperature is reduced so that it can be burnt at the bottom of the convective layer and at the same time Be diffusion is enhanced so that it can be hidden below the solar convective zone.

We admit that the proposed connection between the laboratory and stellar interiors is highly hypothetical and that so far there is no explanation for the enhancement of astrophysical S-factors at low energies, nevertheless this enhancement is now the result of several experiments and the values we derive for the effective plasma screening potential are relatively similar to those measured in the laboratory.

We believe that a series of new experiments and observations is necessary for testing/disproving our approach:
i) the cross sections of p+$^7$Li → $\alpha + \alpha$, p+$^9$Be → $\alpha +^6 Li$ , p+$^9$Be → $d + 2\alpha$ reactions should be measured at still lower energies, so that a precise determination of $U_{lab}^{p+Li}$ and $U_{lab}^{p+Be}$ can be obtained;
ii) elastic (transport) cross sections for p+$^7$Li and p+$^9$Be collisions should be measured in an energy range of astrophysical interest: does the enhanced transparency suggested by inelastic reactions also show up in the elastic channel?
iii) precise measurements of Li and Be abundances in intermediate age and old open clusters would obviously be interesting, to compare the time dependence we propose with richer data set.

## ACKNOWLEDGMENTS

We are very grateful to C. Rolfs for his valuable comments and suggestions and to S. Degl'Innocenti for useful discussions about properties of the convective envelope from evolutionary calculations.

# APPENDIX A: THE EVOLUTIONARY CODE

The stellar models presented in this paper have been computed by means of the latest version of the FRANEC (Frascati Raphson Newton Evolutionary Code), described by Chieffi & Straniero (1989; CS89). Let us recall the relevant physical inputs and computational procedures and comment on some recent improvement.

This version of the code differs from that described in CS89, mainly because the equations of chemical evolution and stellar structure are simultaneously solved. In addition a full network that includes all stable isotopes up to $^{64}$Ni is explicitly included.

The equation of state takes into account the quantum-relativistic effects for the electron component of the stellar plasma and the electrostatic interaction (Straniero 1988).

Tables of radiative opacity coefficients are derived form Iglesias, Roger and Wilson (1992) for T> $10^4$ K and from Kurucz (1991) at lower temperature. The heavy element solar mixture of Grevesse (1991) has been adopted (except for low-temperature opacity for which a slightly different mixture has been used by Kurucz, namely Anders & Grevesse 1989).

Nuclear reaction rates are generally taken from Caughlan & Fowler (1988), whereas for the $^7$Li(p,$\alpha$) reaction we have adopted the more recent cross section measurements by Engstler et al. 1992. Electron screening is derived form Graboske et al. 1973 (see also DeWitt, Graboske and Cooper 1973).

As usual, the mixing length has been calibrated by comparing the theoretical radius of the standard solar models (SSM) with the observed one (see Chieffi, Straniero and Salaris, 1995). In such a way we have obtained $\alpha = 1/H_p$=2.25 for the no-diffusion SSM and $\alpha = 2.40$ when diffusion of H, He, C, N, O, Li and Be are taken into account following Thoule et al. 1994. The other relevant features of our SSM are reported in Table 5.

The allowed variations of r, L, P, T, M(r) and temperature gradient between two adjacent mesh points should not be in excess of some prefixed values, namely: $\delta r/r$=0.1, $\delta L/L$ =0.01, $\delta P/P$=0.05, $\delta T/T$=0.02, $\delta M/M$=0.01 and $\delta \nabla T/\nabla T$=0.1. As a consequence, the typical number of mesh points required for an integration for a PMS or a MS models ranges between 500 and 600. About 600 time-steps are required for a SSM, 400 of which for the PMS.

# APPENDIX B: THE BOTTOM OF THE SOLAR CONVECTIVE ZONE

We present in Table 5 a summary of information about the bottom of the convective zone in the present Sun, as obtained from helioseismological observations and from the results of several standard solar model calculations. Concerning our results (F), these have been obtained by using the FRANEC code with and without diffusion (see Appendix A).

From inversion of the solar data (Christensen-Dalsgaard et al. 1991) it has been possible to derive the depth of the convective zone ($d_b$= 1- $R_b/R_\odot$) and the sound speed, $c_b$. The temperature is then estimated from the sound speed, by using the perfect gas law and assuming the He mass fraction to be in the range Y=0.23-0.29:

$$T_b = (2.2 \pm 0.1) \quad 10^6 K \qquad (34)$$

Concerning standard solar model calculations, when $T_b$ ($c_b$) was not given by the authors, it was calculated by us from the value of $c_b$ ($T_b$) again by using the fully ionized perfect gas law.

As is well known (Bahcall & Pinsonneault 1992; Christensen-Dalsgaard, Proffit and Thompson 1993; Proffit 1994), in models without diffusion the convective zone is thinner and the basis is cooler than given from helioseismological data. On the other hand, models where diffusion is taken into account look closer to the real Sun.

Coming to the quantities which are relevant for diffusion, we note that the gradients $dlnP/d(R/R_\odot)$ and $dlnT/d(R/R_\odot)$ are similar (to the level of 10% or better) among the different models.

We define a P-T diffusion rate parameter as:

$$\left(\frac{dy_i}{dt}\right)_{PT} = log_{10}e \ v_{iPT} \ \rho_i S/M_i \quad , \tag{35}$$

where $v_{iPT}$ is the diffusion velocity due to pressure and thermal gradient, see equation (27). It is important that $(\frac{dy_i}{dt})_{PT}$ does not change by more than 20% when different solar models are used.

The time dependence (along the main sequence) of the physical parameters, $O_\alpha$ at the bottom of the convective zone has been investigated numerically within the FRANEC code (with and without diffusion). As expected all quantities vary smoothly and weakly and are well approximated by a linear function of time:

$$O_\alpha = O_{\alpha\odot}[1 + \beta_\alpha(t - t_\odot)/t_\odot] \quad . \tag{36}$$

In Table 3 the values of $\beta_\alpha$ are shown. We remark that nuclear burning rates are very sensitive to $T_b$. Thus even a 10% variation along the main sequence can be significant.

On the other hand we note that the diffusion rate parameters $(\frac{dy_i}{dt})_{PT}$ are constant in time within 10% or better. From Table 5 one thus derives that the photospheric Li and Be depletions during the solar main sequence are, for $\eta = 5$:

$$\Delta y_{Li}^{PT}(MS) \approx \Delta y_{Be}^{PT}(MS) \approx 0.06 \quad . \tag{37}$$

TABLES

TABLE 1. Data on Be abundances ($y_{Be} = log_{10}(N_{Be}/N_H) + 12$).

|  | $y_{Be}$ | Ref. |
|---|---|---|
| Meteorites | 1.42± 0.04 | Anders & Grevesse (1989) |
| 1$M_\odot$ star in the Hyades | 0.9±0.3 | Garcia Lopez et al. (1994) |
| Solar photosphere | 1.15± 0.10 | Anders & Grevesse (1989) |

TABLE 2. The experimentally determined effective screening potential, $U_{lab}$, together with the theoretical value in the adiabatic limit, $U_{el}$, calculated following equation (13).

| reaction | $U_{lab}[eV]$ | | $U_{el}[eV]$ | Ref. |
|---|---|---|---|---|
|  | atomic target | molecular target | | |
| $d +^3 He \to \alpha + p$ | 186 ± 9 | 123±9 | ≈110 | (a) |
| $p +^6 Li \to \alpha +^3 He$ | 470±150 | 440±150 | ≈180 | (b) |
| $p +^7 Li \to \alpha + \alpha$ | 300±280 | 300±160 | ≈180 | (b) |
| $d +^6 Li \to \alpha + \alpha$ | 380±250 | 330±120 | ≈180 | (b) |
| $p +^9 Be \to d + 2\alpha$ $p +^9 Be \to \alpha +^6 Li$ |  | 0-2000*? | ≈260 | (c) |
| $p +^{11} B \to \alpha +^8 Be$ |  | 430±80 | ≈345 | (d) |

(a) Prati et al. 1994, (b)=Engstler et al 1992, (c)=Tombrello & Sierk 1989, (d)=Angulo et al 1993.

∗ value estimated by us.

TABLE 3. Time dependence for the physical parameters, $O_\alpha$, of the Sun during the main sequence phase, at the bottom of the convective zone, We present the values of the slopes $\beta_\alpha$ in the linear approximation $O_\alpha = O_{\alpha\odot}[1 + \beta_\alpha(t - t_\odot)/t_\odot]$ as calculated through FRANEC without (a) and with (b) diffusion.

|  | (a) | (b) |
|---|---|---|
| $T_b$ | -0.10 | -0.07 |
| $\rho_b$ | -0.75 | -0.70 |
| $R_b$ | 0.1 | 0.1 |
| $M_{conv}$ | -0.25 | -0.13 |
| $dlnP/d(R/R_\odot)$ | -0.15 | -0.16 |
| $dlnT/d(R/R_\odot)$ | -0.12 | -0.18 |
| $(dy_{Li}/dt)_{PT}$ | 0.06 | 0.01 |

TABLE 4. Diffusion coefficients at the bottom of the solar convective zone, calculated for $T_b = 2.2 \cdot 10^6$ K, $\rho = 0.2$ $g/cm^3$ and $X_b = 0.7$, following Thoule et al 1994.

| element | $A_p$ | $A_T$ | $A_c$ |
|---|---|---|---|
| $^7$Li | 0.9 | 1.2 | -0.1 |
| $^9$Be | 0.8 | 1.3 | -0.06 |

TABLE 5. Comparison among different standard solar models without and with diffusion; the helioseismological results (H) are also shown. The labels correspond to: H=Cristensen-Dalsgaard, Gough and Thompson (1991); CPT=Christensen-Dalsgaard, Proffitt and Thompson (1993); P= Proffitt (1994); TCL=Turck-Chièze & Lopes (1993); BP=Bahcall & Pinsonneault (1992). F is for the FRANEC code we are using.

| | H | F | CPT | TCL | P | BP | CPT | F | P | P | BP |
|---|---|---|---|---|---|---|---|---|---|---|---|
| Fixed parameter | | Z | Z | Z/X | Z/X | Z/X | Z | Z | Z/X | Z/X | Z/X |
| Diffusion | | no | no | no | no | no | He | He,Z | He | He,Z | He |
| $L_\odot [10^{33} erg/s]$ | | 3.83 | 3.85 | 3.85 | 3.85 | 3.86 | 3.85 | 3.83 | 3.85 | 3.85 | 3.86 |
| $R_\odot [10^{10} cm]$ | | 6.96 | 6.96 | 6.96 | 6.96 | 6.96 | 6.96 | 6.96 | 6.96 | 6.96 | 6.96 |
| $t_\odot [Gy]$ | | 4.6 | 4.6 | 4.5 | 4.63 | 4.6 | 4.6 | 4.6 | 4.63 | 4.63 | 4.6 |
| $Y^{iniz} \times 10$ | | 2.910 | 2.798 | 2.714 | 2.729 | 2.716 | 2.777 | 2.890 | 2.740 | 2.803 | 2.727 |
| $Z^{iniz} \times 10^2$ | | 2. | 2. | 1.770 | 1.907 | 1.895 | 2. | 2. | 1.988 | 2.127 | 1.958 |
| $(Z/X)^{iniz} \times 10^2$ | | 2.903 | 2.856 | 2.490 | 2.694 | 2.671 | 2.848 | 2.894 | 2.802 | 3.045 | 2.766 |
| $Y^{surf} \times 10$ | | 2.910 | 2.798 | 2.714 | 2.724 | 2.716 | 2.494 | 2.552 | 2.456 | 2.514 | 2.466 |
| $Z^{surf} \times 10^2$ | | 2. | 2. | 1.770 | 1.907 | 1.895 | 2. | 1.849 | 1.979 | 1.964 | 1.958 |
| $(Z/X)^{surf} \times 10^2$ | | 2.903 | 2.856 | 2.490 | 2.694 | 2.671 | 2.737 | 2.546 | 2.694 | 2.694 | 2.668 |
| $(R_b/R_\odot) \times 10$ | 7.13±0.03 | 7.29 | 7.22 | 7.25 | 7.25 | 7.21 | 7.08 | 7.13 | 7.103 | 7.115 | 7.07 |
| $c_b [10^7 cm/s]$ | 2.23±0.02 | 2.14 | 2.23 | 2.23 | | 2.19* | 2.23 | 2.23 | | | 2.25* |
| $T_b [10^6 K]$ | 2.21±0.10 | 2.09 | 2.22* | 2.11 | | 2.13 | 2.22* | 2.21 | | | 2.26 |
| $\rho_b [g/cm^3]$ | | 0.154 | | | | 0.167 | | 0.179 | | | 0.197 |
| $(M_{conv}/M_\odot) \times 10^2$ | | 2.03 | | 2.06 | | 2.16 | | 2.35 | 2.48 | 2.48 | 2.54 |
| $dlnP/d(R/R_\odot)$ | | -13 | | | | -12 | | -13 | | | -12 |
| $dlnT/d(R/R_\odot)$ | | - 4.9 | | | | - 4.6 | | - 4.7 | | | - 5.1 |
| $(dy_{Li}/dt)_{PT} [Gy^{-1}]$ | | - 0.014† | | | | - 0.013† | | - 0.013† | | | - 0.013† |
| $(dy_{Be}/dt)_{PT} [Gy^{-1}]$ | | - 0.014† | | | | - 0.012† | | - 0.012† | | | - 0.012† |

* calculated by us assuming fully ionized perfect gas law.

† calculated by us with $\eta = 1$ and diffusion coefficients of Table 4.

FIGURES

FIG. 1. Observed Li abundances, $y_{Li} = log(N_{Li}/N_H) + 12$, in 1$M_\odot$ stars as a function of age. For the open clusters, the error bars are indicative of the spread in the measured values and/or of the uncertainties on the measured $T_{eff}$. Also shown are the measured value in the solar photosphere and the meteoritic value. The straight line is a linear best fit to all data, but the meteoritic. On the right side we also show the estimated pre-main sequence (PMS) and main seqence (MS) Li depletion ($\Delta y_{Li}$) for the Sun.

FIG. 2. The reactivity $N_A < \sigma v >$ for collisions of Li and Be with protons as a function of temperature. For $^9Be$ we use the expression given by Caughlan & Fowler 1988; for $^6Li$ and $^7Li$ we take the more recent results from Engstler et al. 1992.

FIG. 3. Time dependence of the Li abundance, $y_{Li} = log(N_{Li}/N_H) + 12$, in a 1$M_\odot$ stars with solar chemical composition, calculated for different values of the effective plasma screening potential $U_{pl}^{p+Li}$. Diffusion effects are not taken into account in this calculation. Also shown are the meteoritic value, taken as the initial stellar abundance, and the observational result for 1$M_\odot$ stars in the Hyades.

FIG. 4. Time dependence of the Be abundance, $y_{Be} = log(N_{Be}/N_H) + 12$, in a 1$M_\odot$ stars with solar chemical composition, calculated for different values of the effective plasma screening potential $U_{pl}^{p+Be}$. Diffusion effects are not taken into account in this calculation. Also shown is the meteoritic value, taken as the initial stellar abundance.

FIG. 5. Li abundance, $y_{Li}$, for the Hyades dwarfs as a function of the effective temperature, $T_{eff}$. We show data from Boesgaard & Tripicco (1986), circles; from Duncan & Jones (1983), diamonds; from Cayrel et al. (1984), squares, and from Thorburn et al. (1993), diagonal crosses. The cross on the top right is indicative of the observational errors. The three curves represent the calculated behaviour for the values of $U_{pl}^{p+Li}$ [eV] shown in the figure. All curves are normalized to yield (approximately) the observed Li abundance at $T_{eff}$=6300.

FIG. 6. Be abundance, $y_{Be}$, for the Hyades dwarfs as a function of the effective temperature, $T_{eff}$. We show data from Garcia Lopez et al. (1994), filled circles, and from Boesgaard, Heacox and Conti (1977), open circles. The three curves represent the calculated behaviour for the values of $U_{pl}^{p+Be}$ [eV] shown in the figure. All curves are normalized to yield (approximately) the observed Be abundance at $T_{eff}$=6300.

FIG. 7. Time dependence of the Li abundance from ZAMS to the present Sun, calculated assuming $T_{b\odot} = 2.2 \cdot 10^6$K, $U_{pl}^{p+Li}$=700 eV and $\eta$=1. We show the contributions due to nuclear burning (BUR), pressure and thermal diffusion (PT), concentration driven diffusion (CON), and the total (TOT). We also show the result of the standard ($U_{pl}^{p+Li} = 0$ and $\eta = 1$) calculation (dashed line, STA).

FIG. 8. Time dependence of the Li abundance from ZAMS to the present Sun, calculated assuming $T_{b\odot} = 2.2 \cdot 10^6$K, $U_{pl}^{p+Li}$=700 eV and $\eta$=5. Same notation as in Fig. 7

FIG. 9. Variation of the Be abundance between the Sun age and the ZAMS as a function of $U_{pl}^{p+Be}$, for different values of $T_{b\odot}$, calculated:
a) for $\eta$= 1 (full lines),
b) for $\eta$= 5 (dashed lines)

FIG. 10. Variation of the Li abundance between the Sun age and the ZAMS a function of $U_{pl}^{p+Be}$, for different values of $T_{b\odot}$, calculated:
a) for $\eta$= 1 (full lines),
b) for $\eta$= 5 (dashed lines)

FIG. 11. Time dependence of the Li abundance, $y_{Li} = log(N_{Li}/N_H) + 12$, in a 1M$_\odot$ star with solar chemical composition, calculated for different values of the effective plasma screening potential $U_{pl}^{p+Li}$ in eV. Diffusion effects are taken into account, see text. The multiplicative factor for Li and Be diffusion is set $\eta_{Li,Be}$=5. Also shown are the observational data.

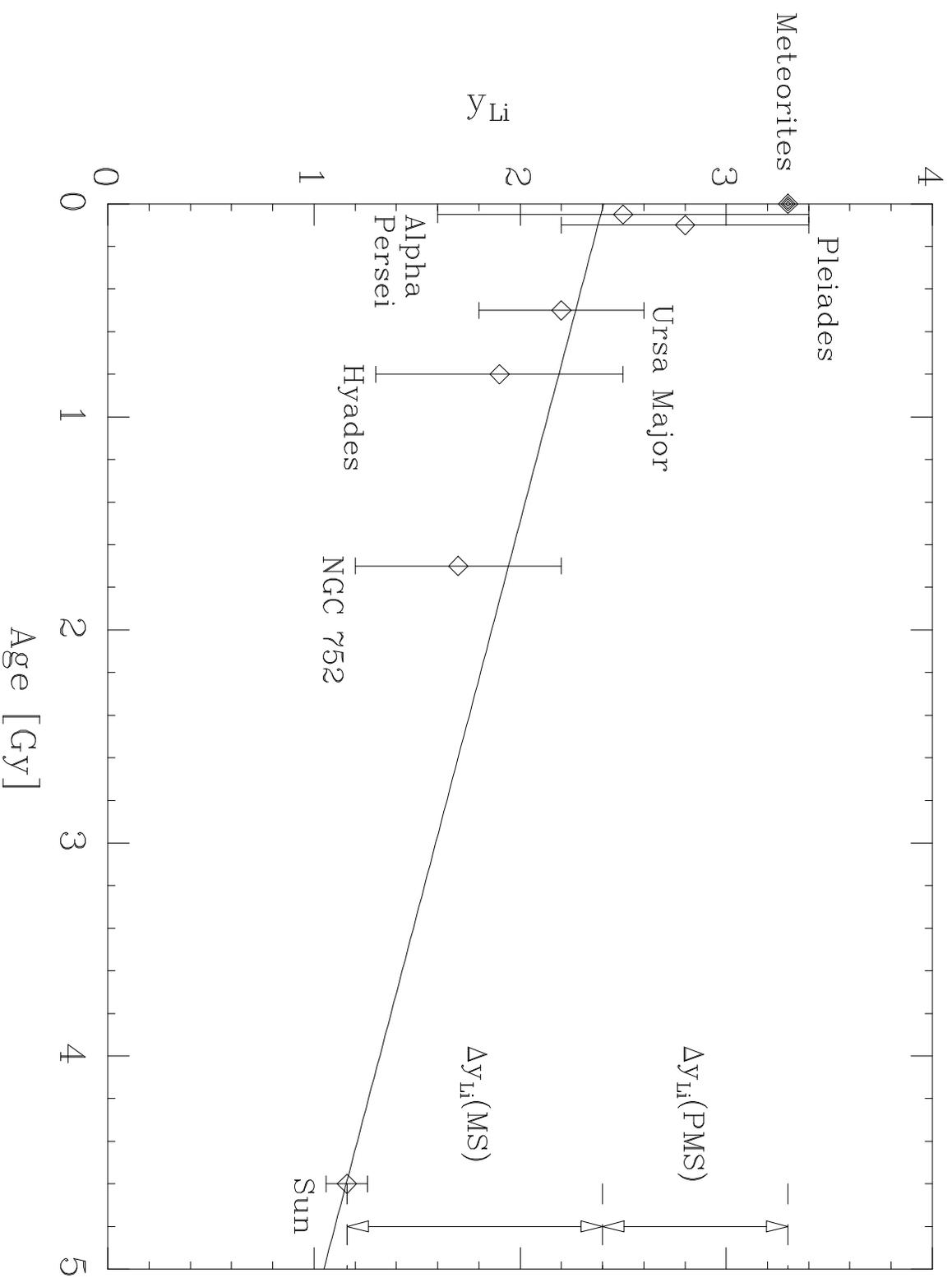

Fig. 1

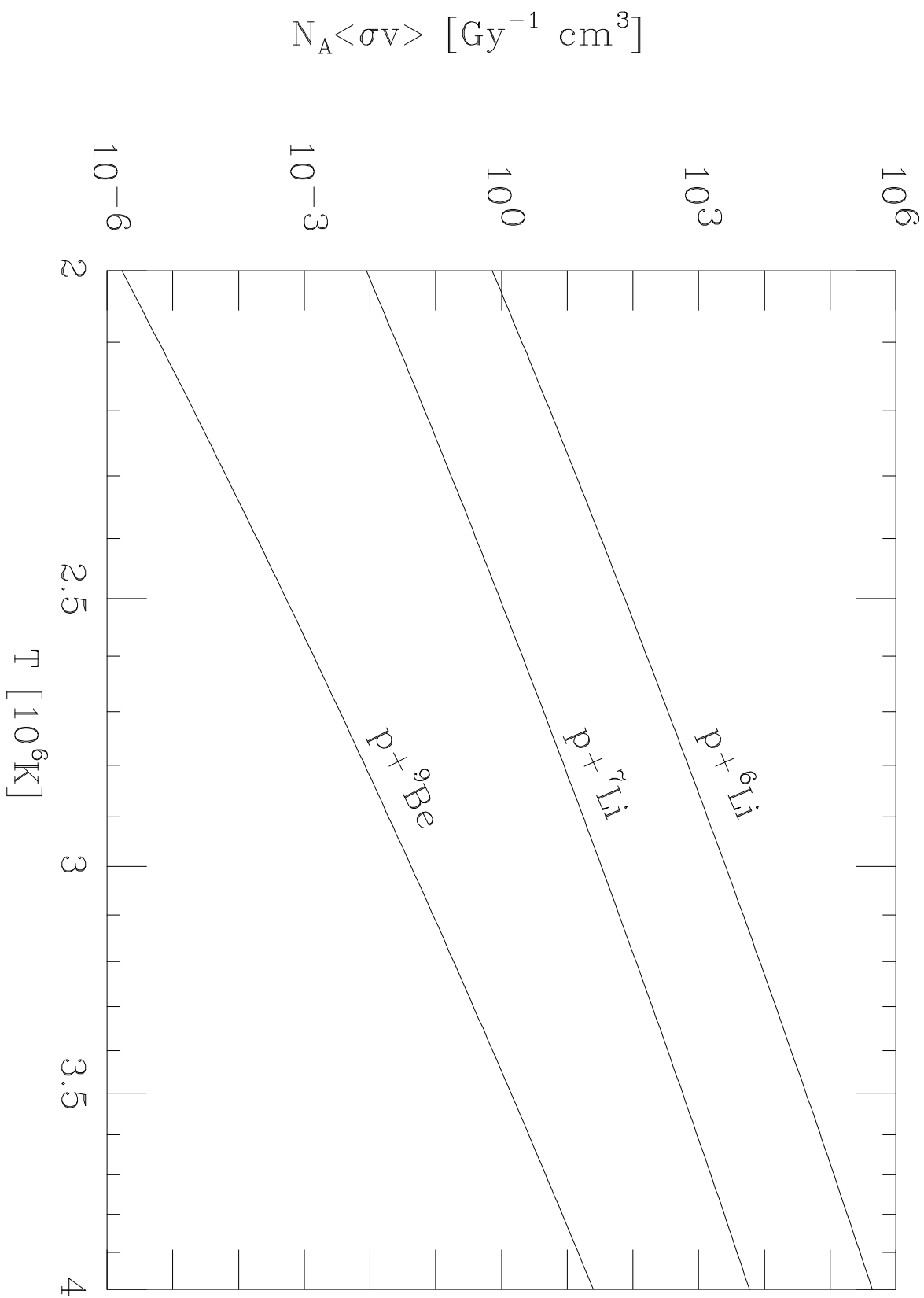

Fig. 2

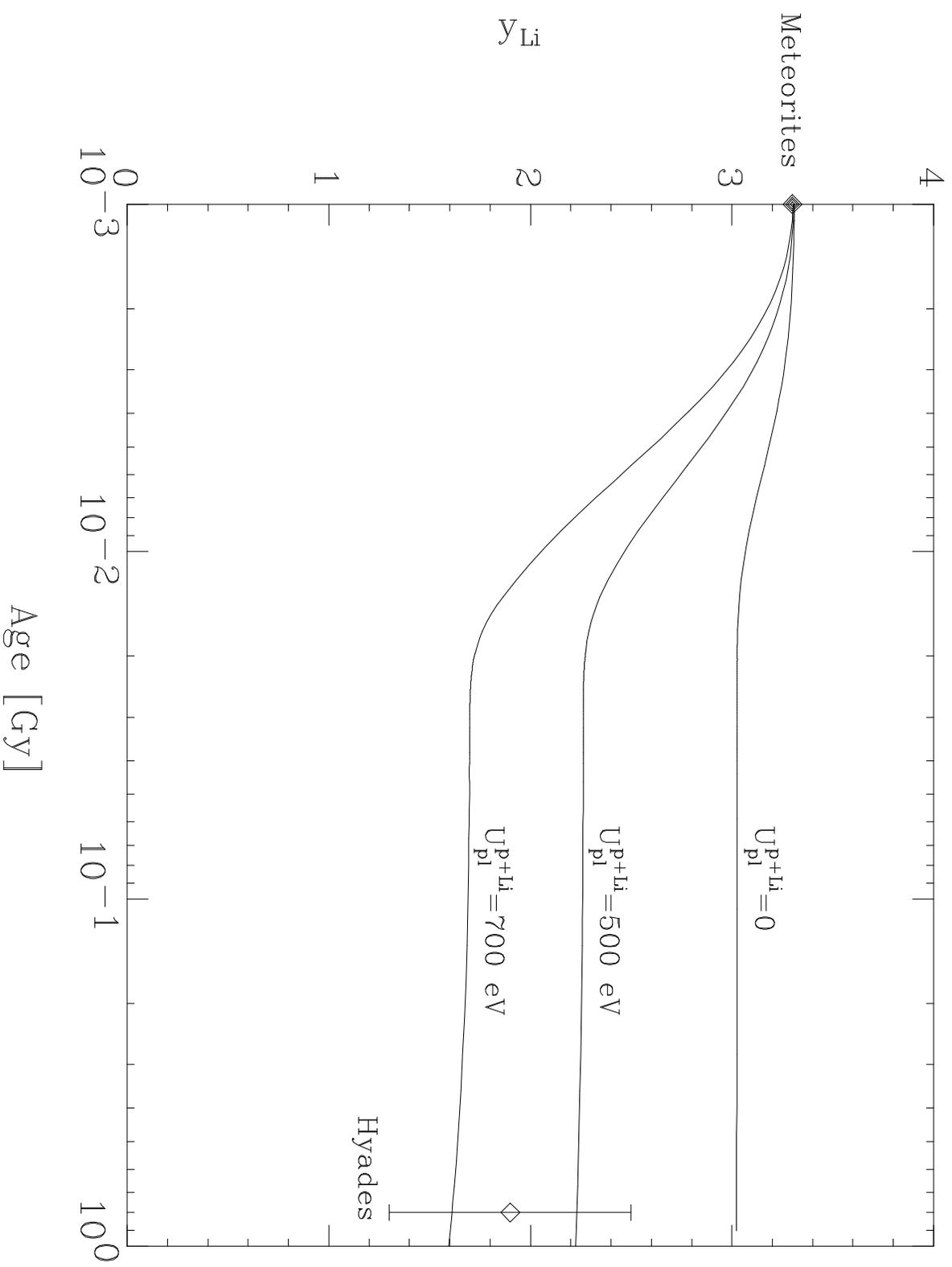
Fig. 3

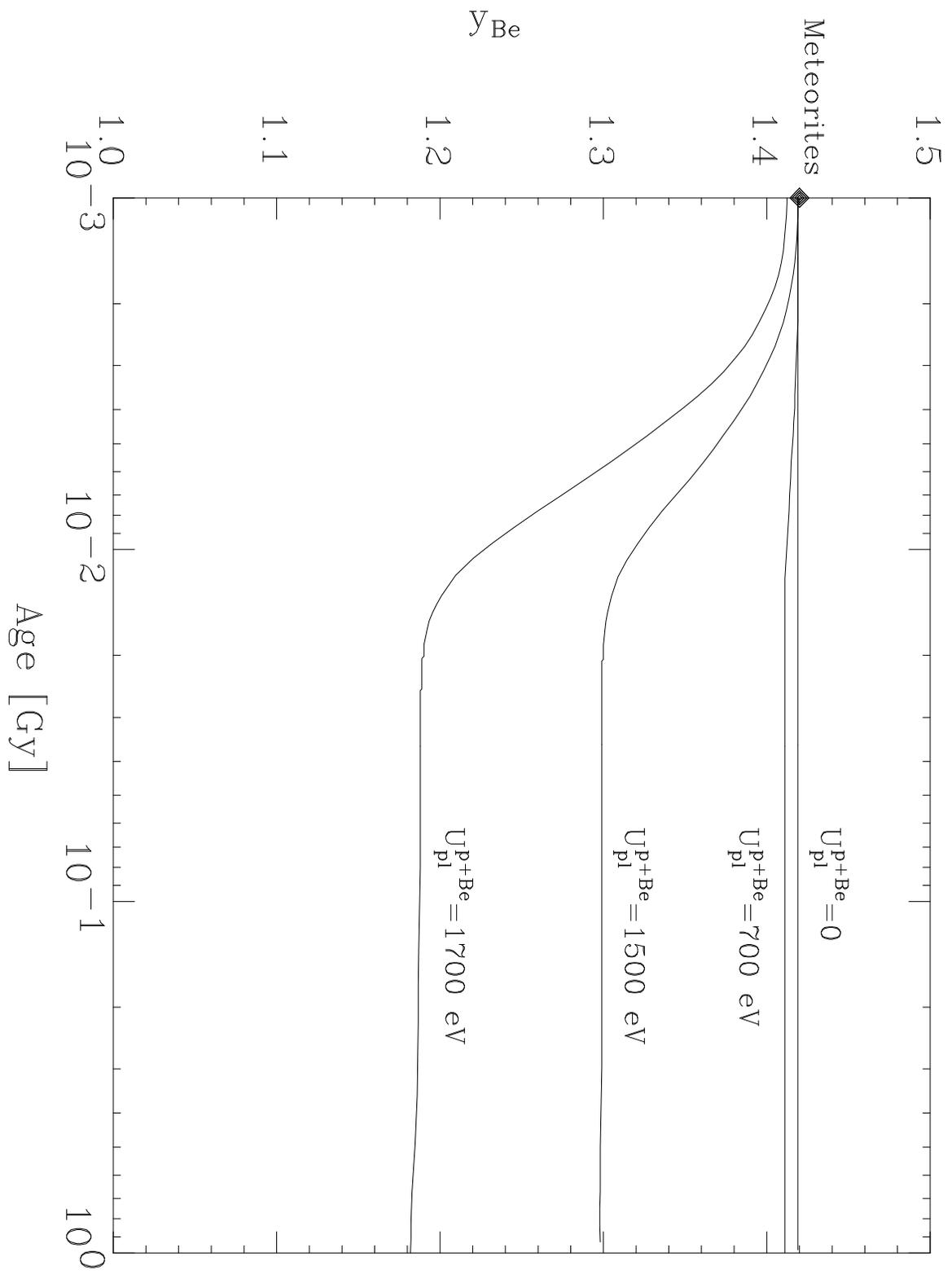

Fig. 4

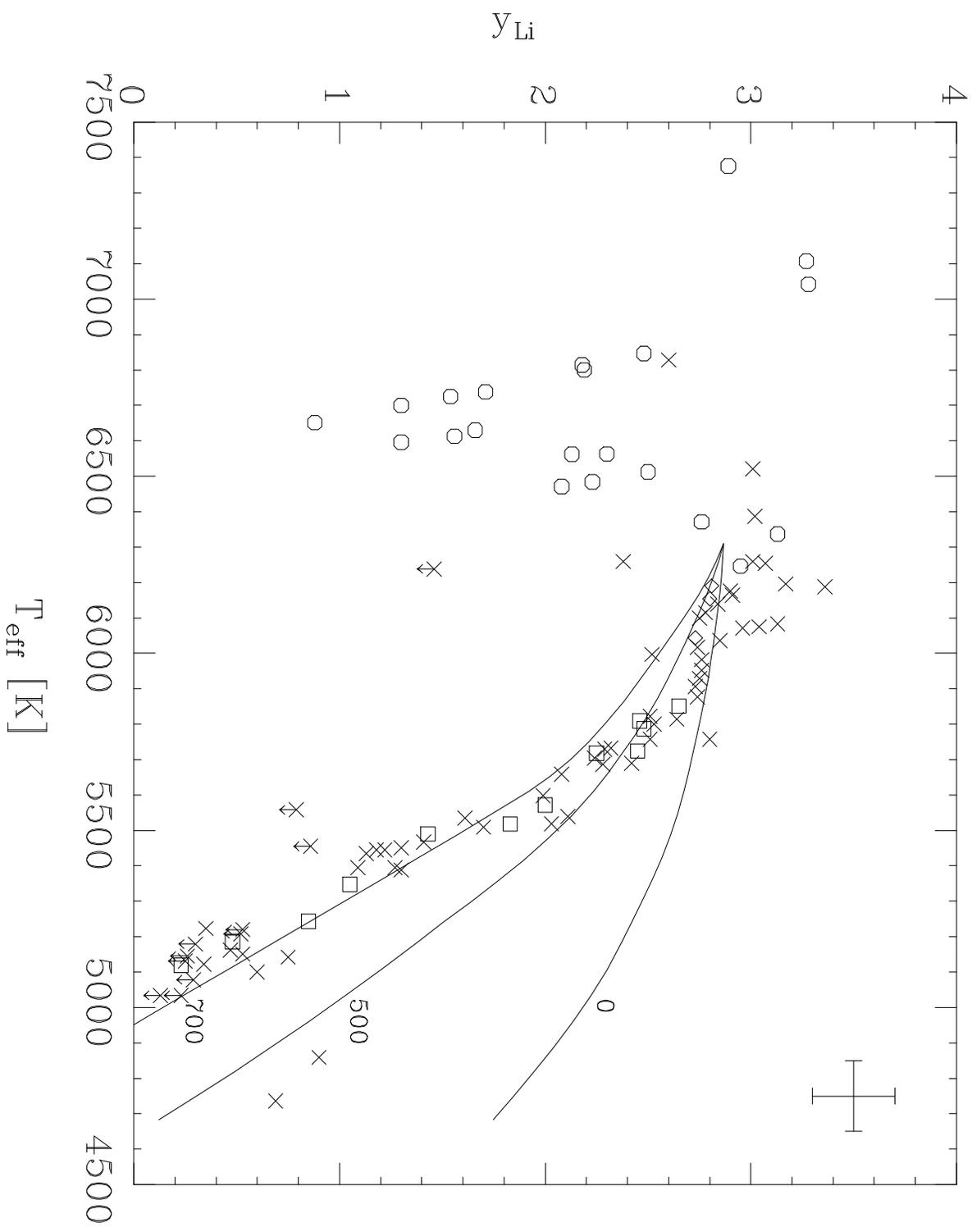

Fig. 5

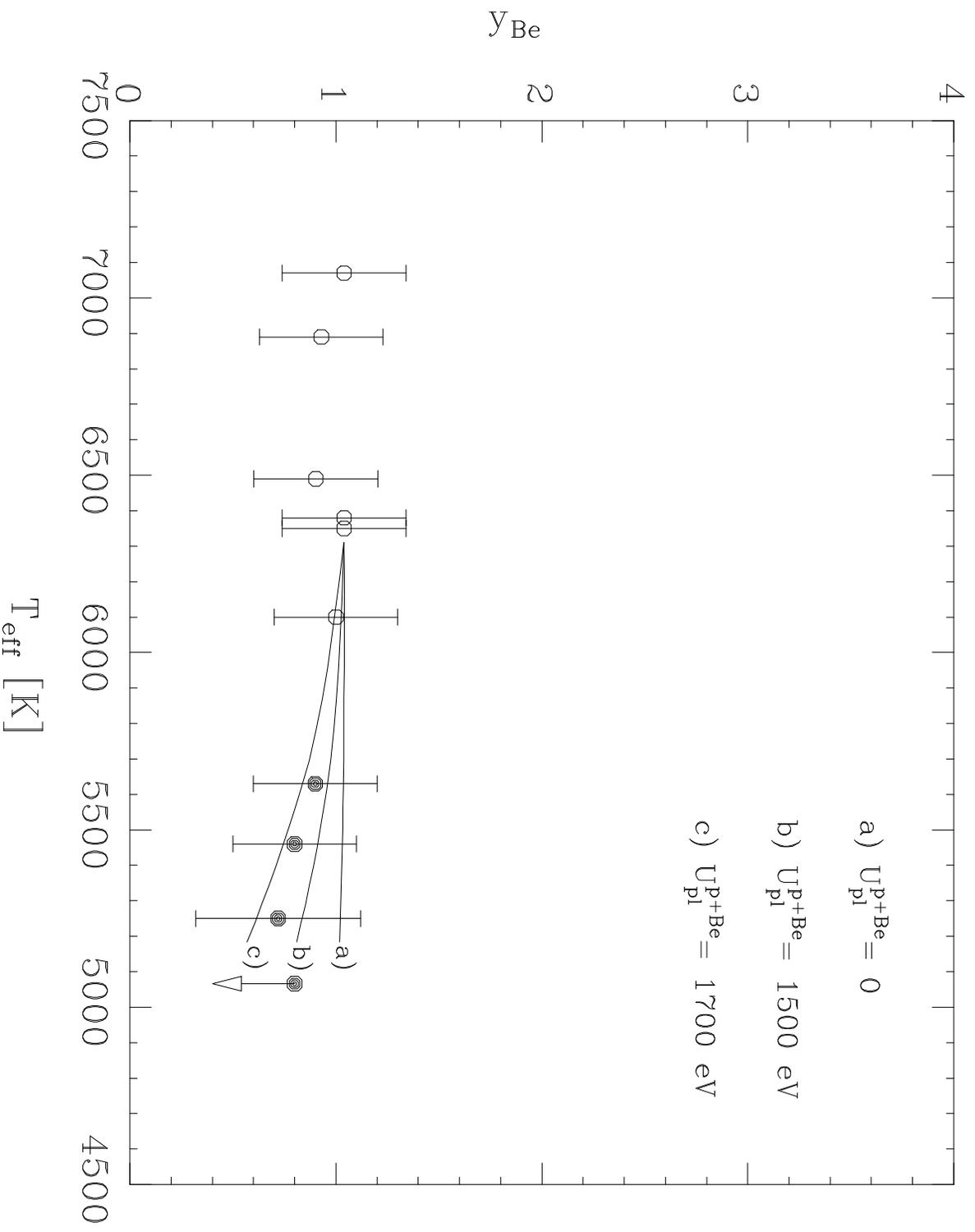

Fig. 6

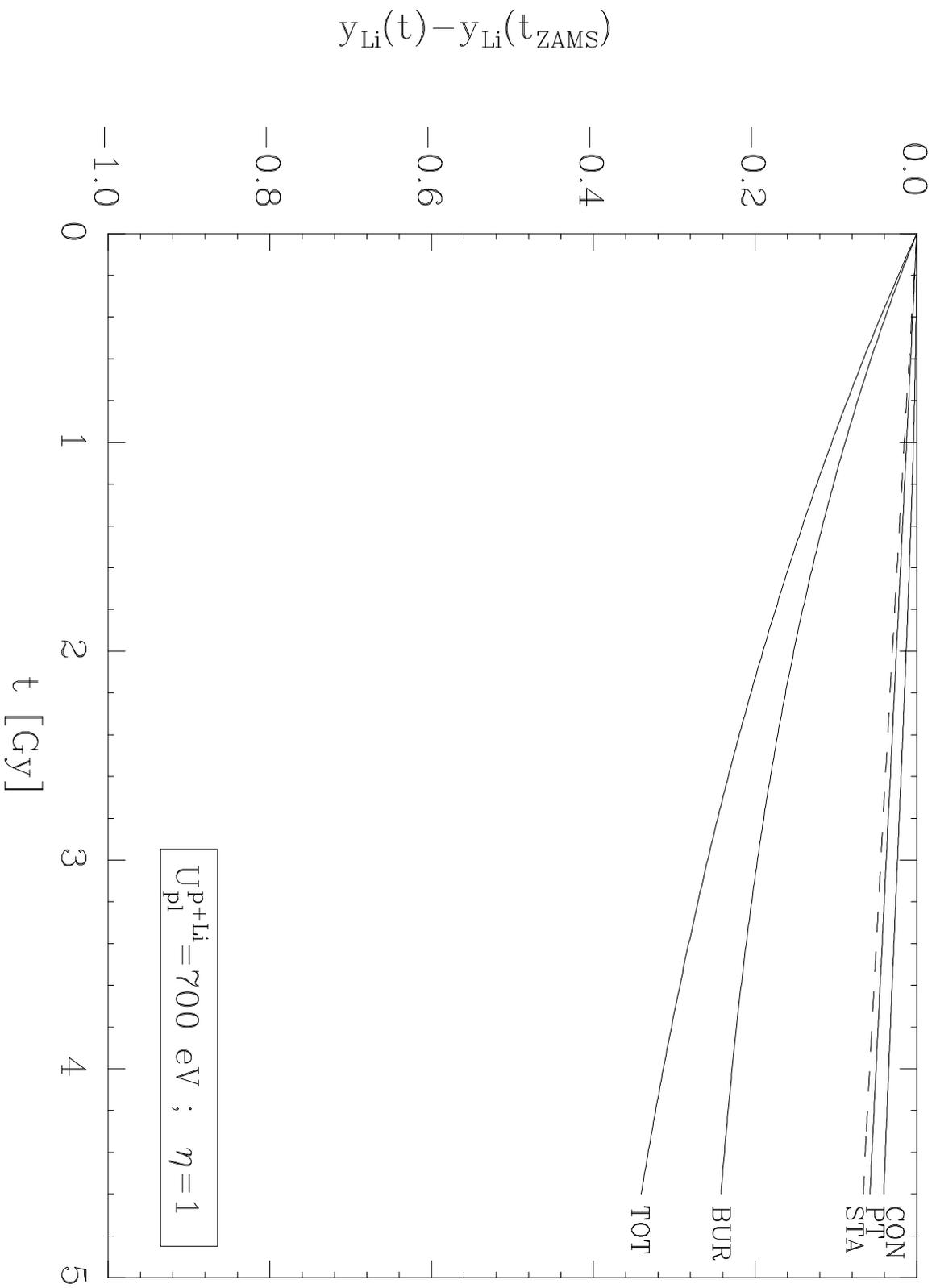

Fig. 7

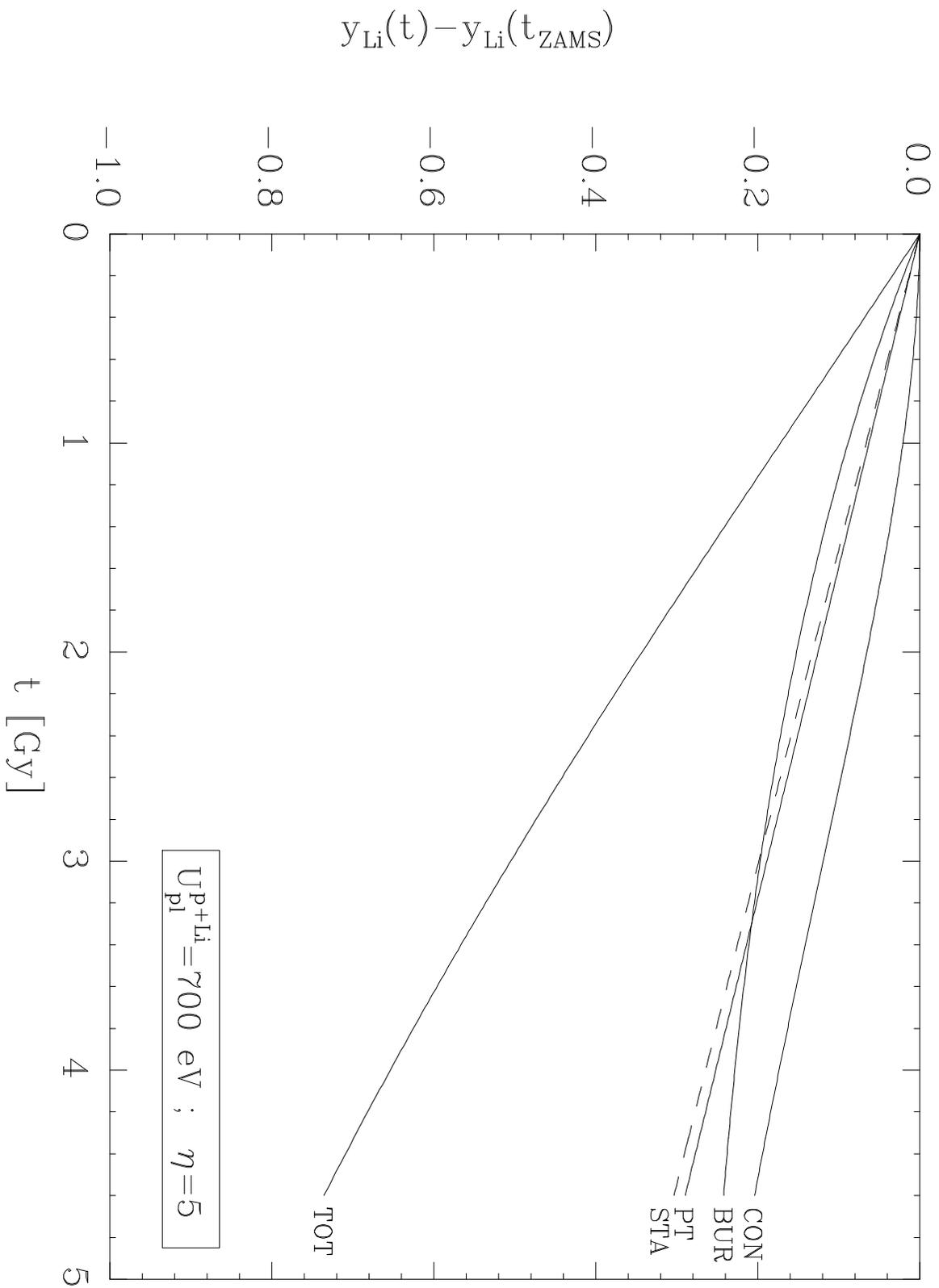

Fig. 8

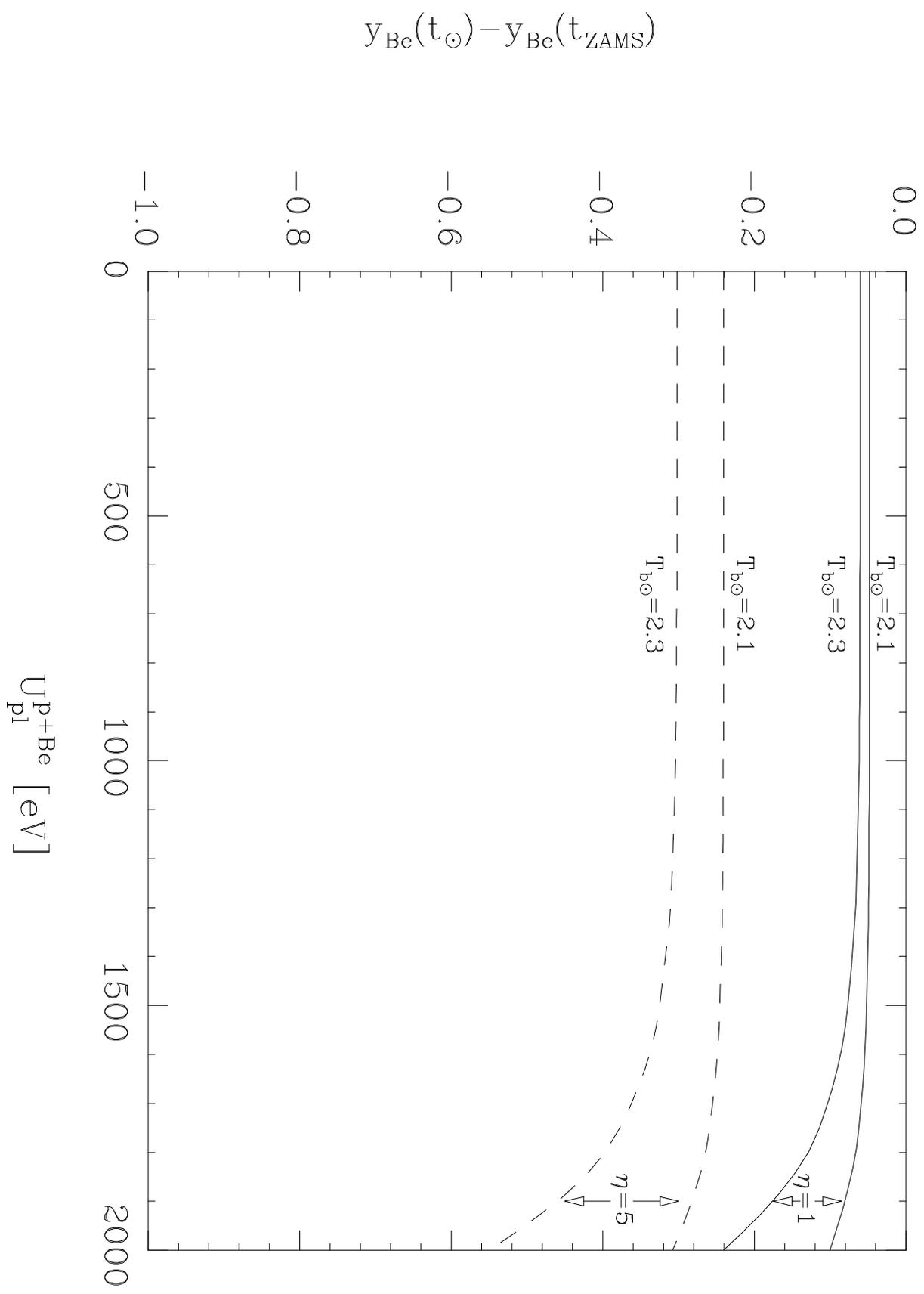

Fig. 9

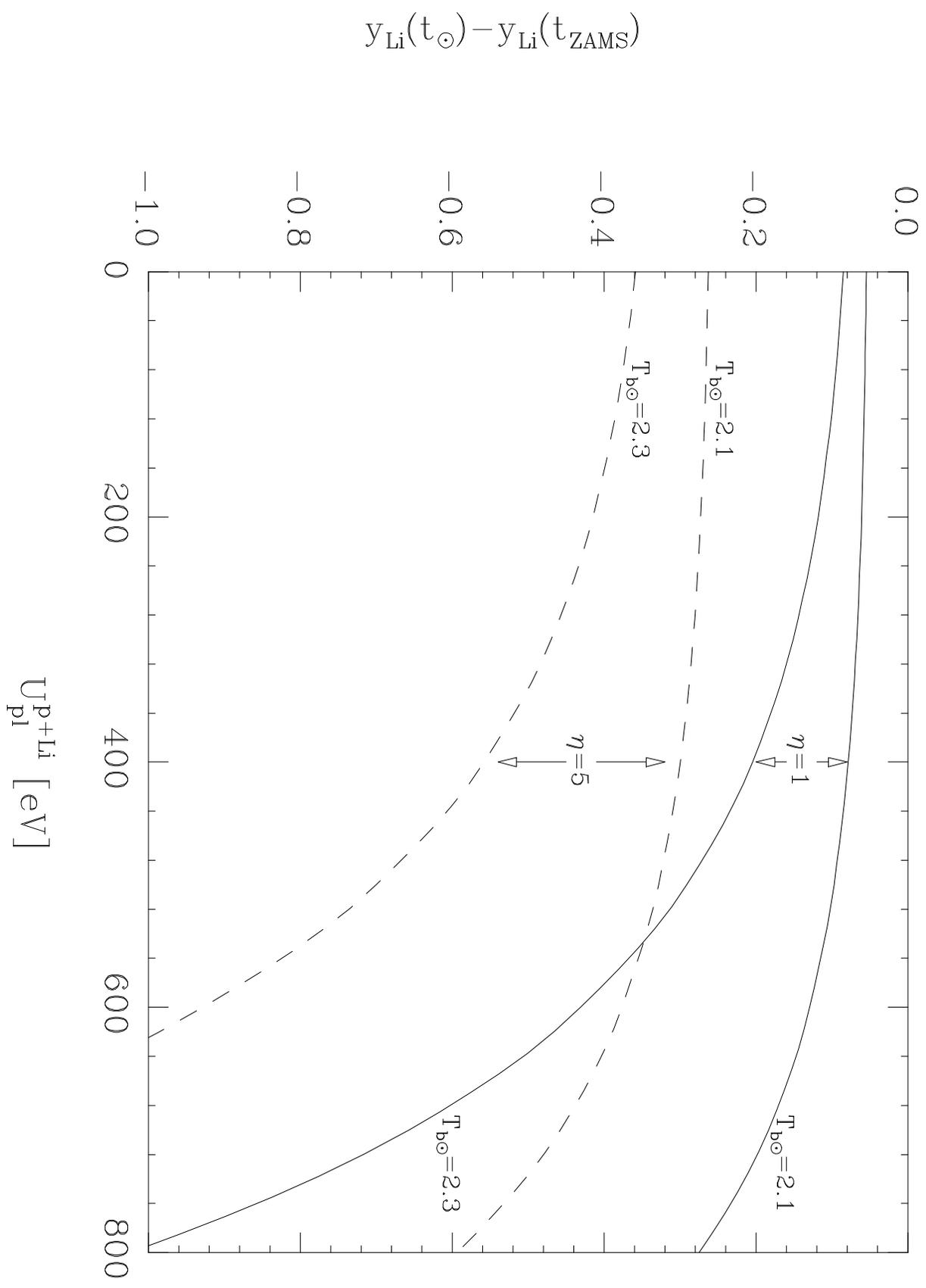

Fig. 10

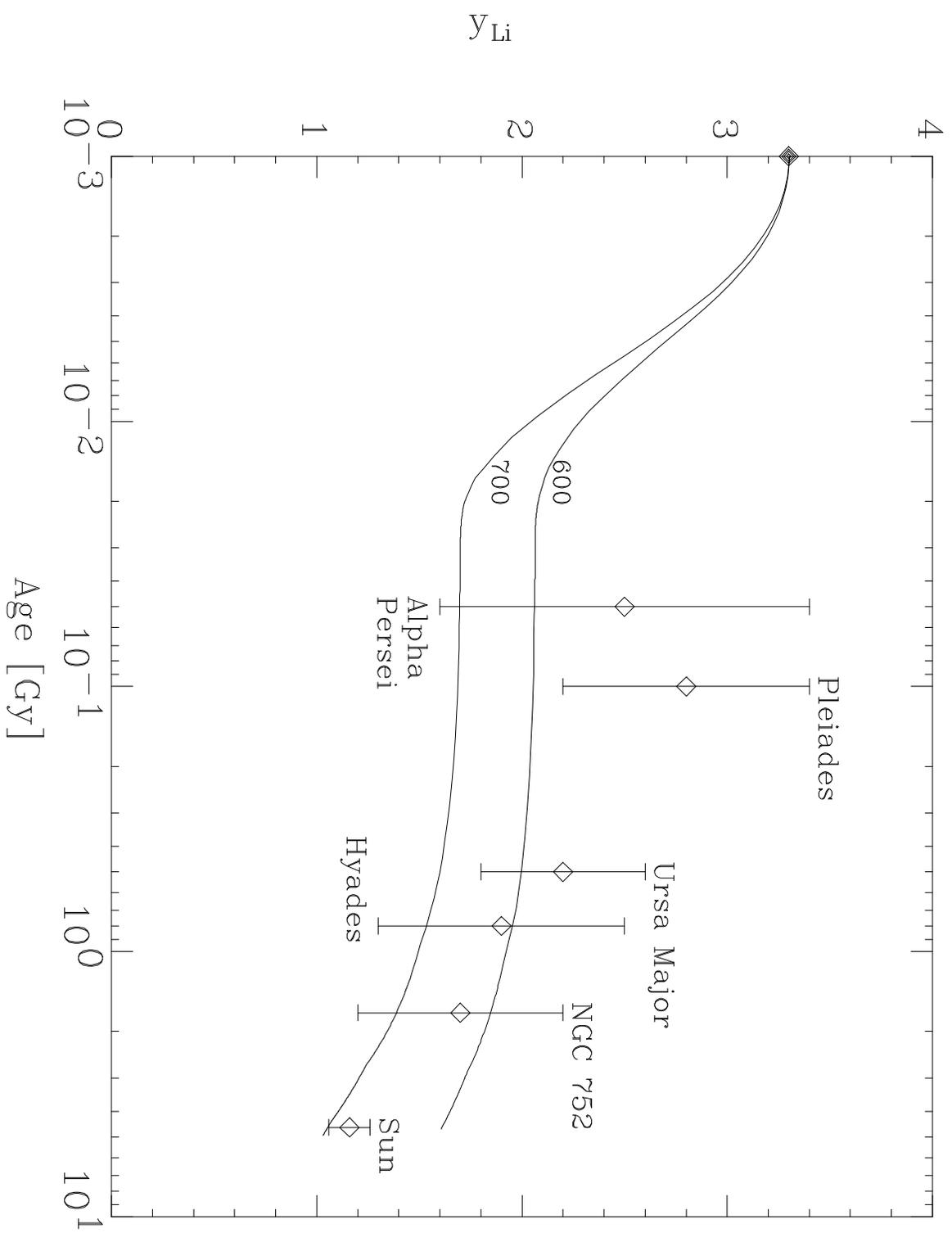

Fig. 11